\documentclass[preprint,11pt]{article}
\usepackage{amsfonts}
\usepackage{color}
\usepackage{graphicx}
\usepackage[dvips]{epsfig}
\usepackage{graphics} % for pdf, bitmapped graphics files
\usepackage{times} % assumes new font selection scheme installed
\usepackage[cmex10]{amsmath} % assumes amsmath package installed
\usepackage{amssymb}  % assumes amsmath package installed
\usepackage{cite}
\usepackage{multirow}
\usepackage[tight,footnotesize]{subfigure}
\usepackage{algorithm}
\usepackage{amsthm}
\usepackage{mathtools}
\usepackage{hyperref}
\usepackage{MnSymbol}

\def\norm #1{\left\|#1\right\|}

\def\twon #1{\left\|#1\right\|_2}

\def\frobn #1{\left\|#1\right\|_{\text{F}}}

\def\atomn #1{\left\|#1\right\|_{\cA}}

\def\abs #1{\left|#1\right|}

\def\st{\text{subject to }}

\def\bC{\mathbb{C}}

\def\bR{\mathbb{R}}

\def\bT{\mathbb{T}}

\def\m #1{\boldsymbol{#1}}

\def\cA{\mathcal{A}}

\def\cT{\mathcal{T}}

\def\bee{\begin{equation}}
\def\ene{\end{equation}}

\def\beq{\begin{eqnarray}}
\def\enq{\end{eqnarray}}
\def\lentwo{\setlength\arraycolsep{2pt}}

\newtheorem{lem}{Lemma}

\newtheorem{thm}{Theorem}
\newtheorem{prop}{Proposition}

\def\equ #1{\begin{equation}#1\end{equation}}
\def\equa #1{\begin{eqnarray}#1\end{eqnarray}}
\def\sbra #1{\left(#1\right)}
\def\mbra #1{\left[#1\right]}
\def\lbra #1{\left\{#1\right\}}
\def\diag #1{\text{diag}#1}
\def\tr #1{\text{tr}#1}

\def\rank #1{\text{rank}#1}
\def\st {\text{ subject to }}

\title{Separation-Free Spectral Super-Resolution via Convex Optimization}
\author{Zai Yang\thanks{School of Mathematics and Statistics, Xi'an Jiaotong University, Xi'an, Shaanxi 710049, China} , Yi-Lin Mo$^*$, Gongguo Tang\thanks{School of Electrical, Computer \& Energy Engineering, University of Colorado, Boulder, CO80309, US} , and Zongben Xu$^*$}
\begin{document}
\maketitle

%%%%%%%%%%%%%%%%%%%%%%%%%%%%%%%%%%%%%%%%%%%%%%%%%%%%%%%%%%%%%%%%%%%%%%%%%%%%%%%%
\begin{abstract}
Atomic norm methods have recently been proposed for spectral super-resolution with flexibility in dealing with missing data and miscellaneous noises. A notorious drawback of these convex optimization methods however is their lower resolution in the high signal-to-noise (SNR) regime as compared to conventional methods such as ESPRIT. In this paper, we devise a simple weighting scheme in existing atomic norm methods and show that the resolution of the resulting convex optimization method can be made arbitrarily high in the absence of noise, achieving the so-called separation-free super-resolution. This is proved by a novel, kernel-free construction of the dual certificate whose existence guarantees exact super-resolution using the proposed method. Numerical results corroborating our analysis are provided.

%Convex optimization (to be specific, atomic norm) methods have recently been proposed for multichannel line spectral estimation. They are demonstrated to be advantageous over previous nonparametric and parametric methods in certain challenging scenarios, e.g., when the number of frequency components is unknown, or when part of the sample data is missing or corrupted by heavy noise. However, they cannot reach the performance limits shared by previous methods in the resolution and the sample size. In this paper, we analytically show that these performance limits can be restored in the noiseless case by incorporating a simple weighting scheme into previous atomic norm methods. As an example, given $L$ channels, it is shown that $K+1$ samples per channel are sufficient to resolve $K\leq L$ generic frequency components whose frequencies are closely located. Our result is proved by a new means of construction of the dual certificate which guarantees exact line spectral estimation using the weighted atomic norm technique.
\end{abstract}

\textbf{Keywords:} Spectral super-resolution, separation-free, weighted atomic norm, convex optimization, dual certificate

\section{Introduction}
The problem of resolving several unknown frequency components from discrete samples of their mixture in the time/spatial domain is known as line spectral estimation or spectral super-resolution \cite{stoica2005spectral,candes2013towards}. Its equivalent forms are also referred to as spectral compressed sensing or infinite-dimensional/off-the-grid/continuous compressed sensing \cite{duarte2013spectral, adcock2015generalized,tang2013compressed, yang2016enhancing}. Spectral super-resolution is a fundamental problem in statistical signal processing and has been studied under various topics such as direction-of-arrival estimation in array processing \cite{yang2018sparse}, beamforming in acoustics \cite{xenaki2015grid}, channel estimation in wireless communications \cite{tsai2018millimeter}, and modal analysis in structural health monitoring \cite{li2018atomic}.

In this paper, we consider the general multichannel spectral super-resolution problem. In particular, we have access to an $N\times L$ data matrix $\m{Y}$ composed of equispaced samples $\lbra{y_{jl}}$ that, in the absence of noise, is given by
\equ{y_{jl} = \sum_{k=1}^K e^{i2\pi(j-1)f_k} s_{kl}, \quad j=1,\dots,N, \; l = 1,\dots,L, \label{eq:model}}
where $L$ is the number of channels, $N$ is the sample size per channel, $i = \sqrt{-1}$, and $s_{kl}$ denotes the unknown complex amplitude of the $k$th frequency $f_k$ in the $l$th channel. In \eqref{eq:model}, the frequencies $\lbra{f_k}$ are normalized by the sampling frequency (at a Nyquist rate) and belong to the unit circle $\bT = \mbra{0,1}$, where $0$ and $1$ are identified. Let $\m{a}\sbra{f} = \mbra{1,e^{i2\pi f},\dots,e^{i2\pi(N-1)f}}^T$ denote a sampled complex sinusoid of frequency $f$ and $\m{s}_k = \mbra{s_{k1},\dots,s_{kL}}$ be the coefficient vector of the $k$th component. The data matrix $\m{Y}$ is then expressed as
\equ{\m{Y} = \sum_{k=1}^K \m{a}\sbra{f_k}\m{s}_k = \m{A}\sbra{\m{f}}\m{S}, \label{eq:model2}}
where $\m{A}\sbra{\m{f}} = \mbra{\m{a}\sbra{f_1},\dots,\m{a}\sbra{f_K}}$ is an $N\times K$ Vandermonde matrix and $\m{S} = \mbra{\m{s}_1^T, \dots, \m{s}_K^T}^T$ is the matrix of coefficients. Given $\m{Y}$ (possibly corrupted by noise), our goal is to recover/estimate $\lbra{f_k}$ and $s_{kl}$.

Spectral super-resolution is complicated by the fact that the data samples are highly nonlinear functions of the frequencies $\lbra{f_k}$ of interest and also by the need of resolving close-located frequencies. To avoid solving a nonconvex optimization problem (e.g., the one induced by the maximum likelihood method), subspace-based methods were developed in the 1970s and since then have dominated the research on this topic for decades. This kind of methods enjoy good statistical properties in the presence of white Gaussian noise \cite{stoica1989music}. They have high resolution in the regime of high signal-to-noise ratio (SNR), and such a resolution can be made arbitrarily high in the limiting noiseless case \cite{liao2016music,yang2022nonasymptotic}. However, their drawbacks are also evident. For example, they need the number $K$ of frequencies that is usually unknown in practice, and their performances deteriorate in the presence of outliers or missing data.

%In practice, an estimate of the covariance matrix is first obtained using the samples of the signal, and then a subspace containing the signal or noise information is retrieved from the estimated covariance matrix, from which the frequencies are finally estimated.

Sparse estimation and compressed sensing methods \cite{candes2006robust, malioutov2005sparse,yang2018sparse}, especially the recent atomic norm approaches \cite{chandrasekaran2012convex, candes2013towards,candes2013super,tang2013compressed, azais2015spike,yang2016exact, li2016off, fernandez2016super,fernandez2017demixing,yang2019sample,li2020approximate}, have been proposed to overcome the aforementioned drawbacks of subspace methods. Differently from the subspace methods, sparse methods use the fact that the unknown number of frequency components is small--the so-called signal sparsity--and attempt to solve for this number jointly with the frequencies by solving a typically convex optimization problem. This optimization framework is flexible in dealing with missing or corrupted samples by taking them as optimization variables. Moreover, differently from earlier compressed sensing methods that only work for discretized frequencies, atomic norm approaches work directly on the continuum, can be cast as semidefinite programming and have provable theoretical guarantees for off-the-grid frequencies.

\begin{figure}[t]
\centering
\includegraphics[width=10cm]{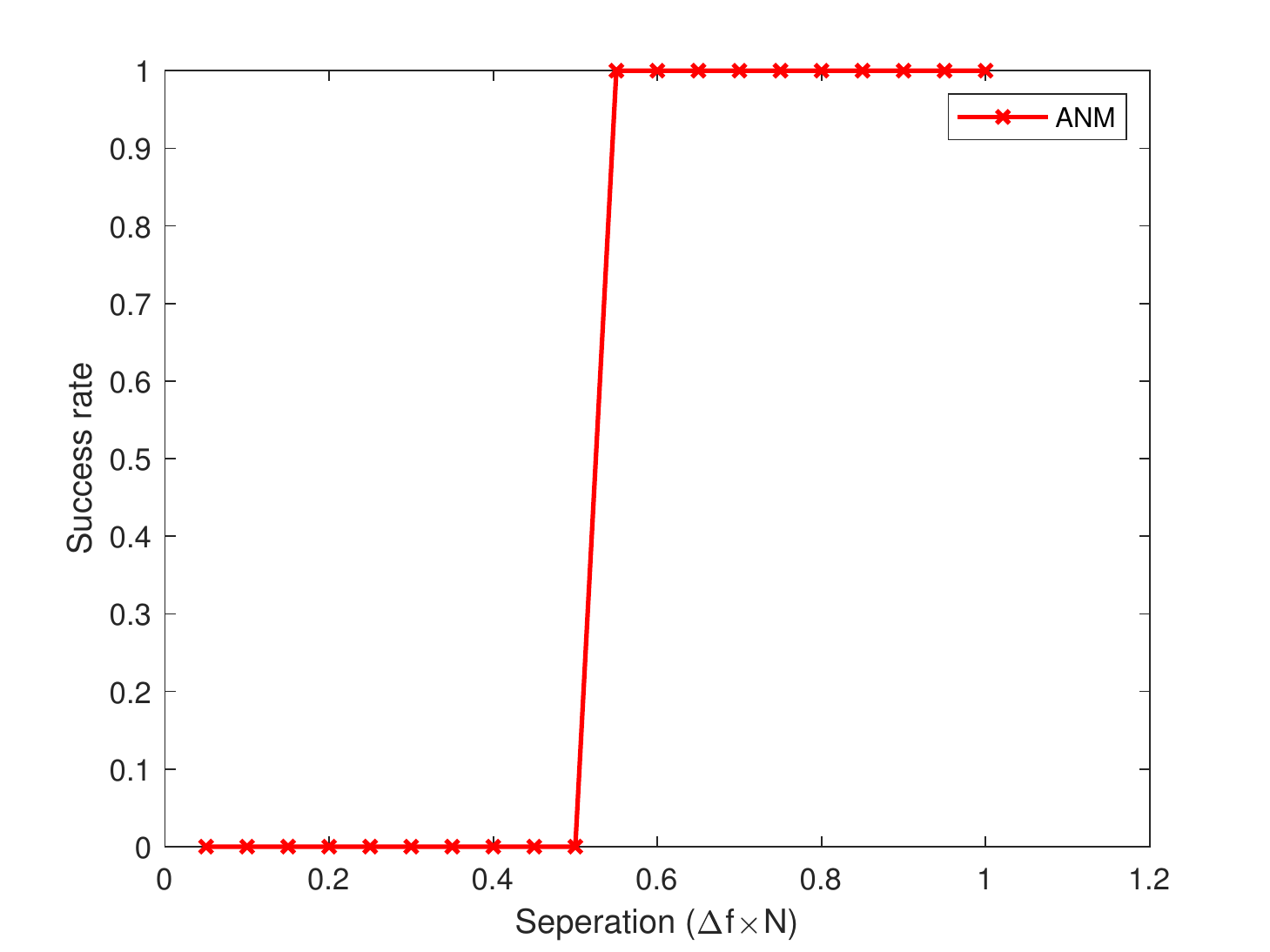}
\caption{Results of success rate of frequency recovery versus frequency separation $\Delta_f$ by using the atomic norm method (ANM) with noiseless data and infinitely many channels, where $N=20$, $K=2$, $f_1=0.1$, and $f_2 = 0.1+\Delta_f$. The frequencies are claimed to be successfully recovered if the root mean squared error is less than $10^{-4}$.}
\label{fig:ANM}
\end{figure}

A key ingredient in the theoretical guarantees for atomic norm approaches is a frequency separation condition (that reflects the resolution limit), to be specific, a wrap-round distance above $4/N$ between any two frequencies. While such a separation condition can be improved using advanced proof techniques (see, e.g., \cite{fernandez2016super}), we provide a simulation result to show that it cannot be removed for atomic norm approaches even in the limiting case with noiseless data and infinitely many channels. In this case, we have access to the noiseless data covariance matrix that under mild assumption is a rank-$K$, positive-semidefinite Hermitian Toeplitz matrix, and the atomic norm method can be implemented equivalently based on this covariance matrix \cite{yang2019sample}. We consider $K=2$ frequency components with unit power and frequencies $\lbra{0.1, 0.1+\Delta_f}$, where $\Delta_f$ varies from $\frac{0.05}{N}$ to $\frac{1}{N}$ and $N=20$. Our numerical results are presented in Fig.~\ref{fig:ANM}. A sharp phase transition is seen due to the fact that the randomness in the sample covariance matrix diminishes as $L\rightarrow+\infty$. The frequencies are exactly recovered as $\Delta_f > \frac{1}{2N}$ and failures occur as $\Delta_f \leq \frac{1}{2N}$.

In contrast to the atomic norm method, it is well-known that subspace methods do not suffer from a resolution limit in the absence of noise. In fact, other conventional methods such as Prony's method or Capon's beamforming \cite{capon1969high} are also known to be separation-free in the noiseless case, though their performances can be sensitive to noise. Therefore, it is crucial to understand and overcome the resolution limit of atomic norm approaches for their application in the high SNR regime. The study of this topic is expected to close the gap between atomic norm and subspace methods as well as to shed light on novel techniques for spectral super-resolution.

We note that separation-free, convex optimization techniques have recently been proposed in literature; see \cite{morgenshtern2016super,bendory2017robust,fuchs2005sparsity, schiebinger2018superresolution, kurmanbek2022multivariate}. However, all these studies assume positive (as opposed to complex in our case) coefficients of the spectral components and simplify the problem at hand.

In this paper, we show that the frequency separation condition can be removed in the atomic norm method if it is modified appropriately. In particular, inspired by \cite{yang2016enhancing}, we devise a weighted atomic norm technique in which a weighting function is constructed from the data and used to assign preference of the frequency candidates in the atomic norm. The weighted atomic norm can be cast as convex programming like the standard atomic norm and it degenerates into the standard one in the case of a non-informative, constant weighting function. We provide a rigorous analysis showing that the frequency components can be exactly recovered by the weighted atomic norm method under mild conditions, without any frequency separation condition. Numerical results are provided that validate our analysis.

Our technical proof is based on a novel construction of the dual certificate which verifies optimality of the true frequencies among all feasible solutions to the proposed convex optimization problem. While such a dual certificate was constructed for the standard atomic norm based on low-pass kernels (see \cite{candes2013towards,fernandez2016super}), we do not use any kernels in our proof. This allows us not to sacrifice the resolution due to the choice of a specific kernel.

Notations used in this paper are as follows. The sets of real and complex numbers are denoted by $\bR$ and $\bC$ respectively. Bold-face letters are reserved for vectors and matrices. For vector $\m{x}$, $\diag\sbra{\m{x}}$ denotes a diagonal matrix with $\m{x}$ on the diagonal. For matrix $\m{X}$, $\diag\sbra{\m{X}}$ forms the diagonal of $\m{X}$ as a column vector. For matrix $ \boldsymbol{A}$, its conjugate, transpose, conjugate transpose, inverse, pseudo-inverse, rank, trace, spectral norm and Frobenius norm are denoted by $\overline{ \boldsymbol{A}}$, $ \boldsymbol{A}^{T} $, $ \boldsymbol{A}^{H} $, $ \boldsymbol{A}^{-1} $, $ \boldsymbol{A}^{\dag} $, $\rank\sbra{\m{A}}$, $\tr\sbra{\m{A}}$, $\twon{\m{A}}$ and $\frobn{\m{A}}$, respectively. That $\boldsymbol{A} $ is Hermitian positive semidefinite is expressed as $\boldsymbol{A}\succcurlyeq 0 $. For matrices $\m{A}$, $\m{B}$ of proper dimensions, their Kronecker, Khatri-Rao (or column-wise Kronecker) and Hadamard products are denoted by $\m{A}\otimes \m{B}$, $\m{A}* \m{B}$ and $\m{A}\odot \m{B}$, respectively. The $N\times N$ identity matrix is denoted by $\m{I}_N$, in which the subscript can be omitted. For function $a(f)$ of $f$, its first and second derivatives are denoted by $\frac{\text{d}a(f)}{\text{d}f}$ and $\frac{\text{d}^2a(f)}{\text{d}f^2}$, or $a'(f)$ and $a''(f)$, or $a^{(1)}(f)$ and $a^{(2)}(f)$ for convenience. The real part of $x$ is denoted by $\Re x$.

The rest of the paper is organized as follows. Section \ref{sec:mainresults} presents our main results consisting of the weighted atomic norm method and its theoretical guarantee. Section \ref{sec:proof} presents the proof of the main theorem. Section \ref{sec:simulation} presents numerical results validating our analysis, and Section \ref{sec:conclusion} concludes this paper.

\section{Main Results} \label{sec:mainresults}

\subsection{Atomic Norm Method}
It is seen from \eqref{eq:model2} that
\equ{\m{Y} = \sum_{k=1}^K \twon{\m{s}_k} \m{a}\sbra{f_k}\frac{\m{s}_k}{\twon{\m{s}_k}}, \label{eq:model3}}
can be written as a mixture of $K$ atoms in the set of atoms
\equ{\cA = \lbra{\m{a}(f)\m{\phi}: \; f\in\bT,\; \m{\phi}\in\bC^{L\times1},\; \twon{\m{\phi}}=1}. \label{eq:atomset0}}
To seek for the spectral components of $\m{Y}$ by exploiting the spectral sparsity in the sense that $K$ is small, the atomic norm of $\m{Y}$ induced by the set of atoms $\cA$ is defined as \cite{chandrasekaran2012convex,yang2016exact,li2016off}:
\equ{\begin{split}\norm{\m{Y}}_{\cA}
=\inf\lbra{\sum_k c_k:\; \m{Y}=\sum_{k} c_k\m{a}_k,\; c_k>0,\; \m{a}_k\in\cA}.\end{split}\label{eq:atomn}}
It is shown in \cite{yang2016exact,li2016off} that the atomic norm $\norm{\m{Y}}_{\cA} $ can be computed using the following semidefinite programming (SDP):
\equ{\norm{\m{Y}}_{\cA} = \min_{\m{X},\m{T}} \frac{1}{2}\tr\sbra{\m{X}}+\frac{1}{2N}\tr\sbra{\m{T}}, \st \begin{bmatrix}\m{X} & \m{Y}^{H} \\ \m{Y} & \m{T}\end{bmatrix} \succcurlyeq \m{0} \text{ and } \m{T} \text{ is Toeplitz}.\label{eq:AN}}
Once \eqref{eq:AN} is solved numerically, we apply the Carath{\'e}odory-Fej{\'e}r theorem \cite[Theorem 11.5]{yang2018sparse} to the solution $\m{T}^*$ and obtain the Vandermonde decomposition (that is unique if $\m{T}^*$ is rank deficient):
\equ{\m{T}^* = \sum_{k=1}^{K^*} p_k^*\m{a}\sbra{f_k^*}\m{a}^H\sbra{f_k^*}= \m{A}(\m{f}^*)\diag\sbra{\m{p}^*}\m{A}^H(\m{f}^*),\quad K^* =\rank\sbra{\m{T}^*}, \label{eq:Vandec}}
where $K^*, \m{f}^*$ are estimates of $K$ and $\m{f}$, respectively.
By using the column inclusion lemma for positive semidefinite matrices, there exist matrix $\m{S}^*$ such that
\equ{\m{Y} = \m{A}(\m{f}^*)\m{S}^* = \sum_{k=1}^{K^*}\m{a}\sbra{f_k^*}\m{s}_k^*, \label{eq:atomdec}}
where $\m{s}_k^*$ denotes the $k$th row of $\m{S}^*$. The atomic decomposition of $\m{Y}$ resulting from \eqref{eq:atomdec} achieves the atomic norm in the sense that $\sum_{k=1}^{K^*}\twon{\m{s}_k^*} = \atomn{\m{Y}}$ \cite{yang2016exact}. Moreover, the above process recovers exactly $\m{f},\m{S}$ if the frequencies are mutually separated by at least $\frac{4}{N}$, an assumption that can be weakened but not removed as shown in Fig.~\ref{fig:ANM}.

\subsection{Weighted Atomic Norm Method}
Suppose that $w(f)>0$ is a weighting function. The weighted atomic norm of $\m{Y}$ associated with $\cA$ and $w(f)$ is defined as \cite{mishra2015spectral,yang2016enhancing}:
\equ{\begin{split}\norm{\m{Y}}_{\cA^w}
=\inf\lbra{\sum_k w(f_k)c_k:\; \m{Y}=\sum_{k} c_k\m{a}_k,\; c_k>0,\; \m{a}_k\in\cA}.\end{split}\label{eq:watomn}}
Intuitively, the smaller the weighting function $w(f)$ is, the more likely the associated frequency $f$ is selected by the weighted atomic norm. In the case when $w(f)$ is constant, the weighted atomic norm degenerates into the atomic norm. Therefore, the key to a better performance by using the weighted atomic norm is to design a good weighting function.

Let
\equ{w(f) =\sqrt{\m{a}^{H}(f)\m{W}\m{a}(f)}, \quad \m{W}\succcurlyeq\m{0}. \label{eq:wf}}
It is shown in \cite{yang2016enhancing} that the weighted atomic norm $\norm{\m{Y}}_{\cA^w}$ can be cast as the following SDP:
\equ{\norm{\m{Y}}_{\cA^w} = \min_{\m{X},\m{T}} \frac{1}{2}\tr\sbra{\m{X}}+\frac{1}{2}\tr\sbra{\m{W}\m{T}}, \st \begin{bmatrix}\m{X} & \m{Y}^{H} \\ \m{Y} & \m{T}\end{bmatrix} \succcurlyeq \m{0}  \text{ and } \m{T} \text{ is Toeplitz}. \label{eq:WAN}}
Evidently, \eqref{eq:WAN} degenerates into \eqref{eq:AN} if we let $\m{W} = \frac{1}{N}\m{I}_N$ and $w(f) \equiv 1$. Once the SDP in \eqref{eq:WAN} is solved numerically, as in the case of the atomic norm, the atomic decomposition of $\m{Y}$ that achieves the weighted atomic norm can be obtained from the Vandermonde decomposition of the solution to $\m{T}$.

In this paper, we use the following weight inspired by \cite{yang2016enhancing,yang2018fast}:
%\equ{\m{W} = \sbra{\m{I} + \epsilon^{-1}\widehat{\m{R}}}^{-1}, \quad \widehat{\m{R}} = \frac{1}{L}\m{Y}\m{Y}^{H},}
\equ{\m{W} = \left\{\begin{array}{ll} \sbra{\m{I} + \epsilon^{-1}\widehat{\m{R}}}^{-1}, & \text{ if $\widehat{\m{R}}$ is rank deficient;} \\ \sbra{\epsilon\m{I} + \widehat{\m{R}}}^{-1}, & \text{ otherwise},  \end{array}\right.}
where $\widehat{\m{R}} = \frac{1}{L}\m{Y}\m{Y}^{H}$ is the sample covariance matrix and $\epsilon>0$ is a regularization parameter. Note that the two choices of the weight differ only by a global scaling factor and thus result in the same frequency estimates. They are made different for computational convenience. By using the second expression of $\m{W}$, we have
\equ{w^{-2}(f) = \frac{1}{\m{a}^{H}(f)\sbra{\epsilon\m{I} + \widehat{\m{R}}}^{-1}\m{a}(f)}, \label{eq:capon}}
which coincides with Capon's beamforming \cite{capon1969high} if $\epsilon$ is omitted. Since the spectrum of Capon's beamforming is usually peaked around the true frequencies, the weighted atomic norm is expected to have good performance, but no rigourous analysis has been derived.

%It is also worth noting that as $\epsilon$ approaches infinity, the weighted atomic norm degenerates into the standard atomic norm.

\subsection{Separation-Free Theoretical Guarantee}
In this paper, we mainly consider noiseless measurements and in this case $\widehat{\m{R}}$ is rank deficient. Our main result is provided in the following theorem.

%Though in this case $\m{W}$ can be explicitly computed as $\epsilon\rightarrow 0$ and inserted into the SDP, this results in numerical instability to the SDP solving. For numerical stability concerns, it is of crucial importance to study the case of $\epsilon>0$.

\begin{thm} Assume we have the $N\times L$ data matrix $\m{Y}=\m{A}\sbra{\m{f}}\m{S}$, where the $K<N$ frequencies $\lbra{f_k}$ are distinct and the $K\times L$ matrix $\m{S}$ has full row rank. Let $\widehat{\m{R}} = \frac{1}{L}\m{Y}\m{Y}^H$ and $\m{W} = \sbra{ \m{I} + \epsilon^{-1}\widehat{\m{R}}}^{-1}$. Then, there exists $\epsilon_0>0$ such that for every $0<\epsilon\leq \epsilon_0$, computing the weighted atomic norm of $\m{Y}$ by solving \eqref{eq:WAN} produces the true frequencies $\lbra{f_k}$. \label{thm:noiseless}
\end{thm}

Theorem \ref{thm:noiseless} states that the frequencies $\lbra{f_k}$ can be exactly recovered by using the proposed weighted atomic norm method if $\epsilon$ is appropriately small. In contrast to the standard atomic norm method, no frequency separation condition is required in Theorem \ref{thm:noiseless}. According to our proof of Theorem \ref{thm:noiseless}, which will be provided in the ensuing section, smaller separations among the frequencies $\lbra{f_k}$ result in more ill-conditioned matrix $\m{A}\sbra{\m{f}}$ and a smaller $\epsilon_0$. This causes potential numerical instability to the SDP solving. In fact, similar numerical issues stem from the problem setup and are also encountered for other super-resolution algorithms, e.g., ESPRIT.

The assumption that $\m{S}$ has full row rank is satisfied if $L\geq K$ and the rows of $\m{S}$ are at general positions, which in fact is also required for subspace methods or Capon's beamforming. In the case when this assumption is not satisfied, e.g., due to limited number of channels (a.k.a., $L<K$) or presence of proportional rows of $\m{S}$, spatial smoothing techniques have been widely studied and successfully employed to restore the performance of subspace methods by creating more, overlapping and shortened measurement vectors from the original measurements $\m{Y}$ \cite{shan1985spatial, yang2022nonasymptotic}. Such spatial smoothing techniques can also be adopted in the proposed weighted atomic norm method.

%
%\begin{cor} A weighted atomic norm method can be devised to recover the frequencies  Assume the $N\times L$ matrix $\m{Y}=\m{A}\sbra{\m{f}}\m{S}$, where the $K$ frequencies $\lbra{f_k}$ are distinct and the $K\times L$ matrix $\m{S}$ has full row rank. Let $\widehat{\m{R}} = \frac{1}{L}\m{Y}\m{Y}^H$ and $\m{W} = \sbra{ \m{I} + \epsilon^{-1}\widehat{\m{R}}}^{-1}$, where $\epsilon >0$. Then, there exists $\epsilon_0>0$ such that for every $0<\epsilon\leq \epsilon_0$, computing the weighted atomic norm of $\m{Y}$ by solving \eqref{eq:WAN} produces the true frequencies $\lbra{f_k}$. \label{thm:noiseless}
%\end{cor}

\section{Proof of Theorem \ref{thm:noiseless}} \label{sec:proof}
We prove Theorem \ref{thm:noiseless} in this section. Our proof follows the routines for atomic norm methods (see \cite{candes2013towards,tang2013compressed, yang2016exact, fernandez2016super,fernandez2017demixing,yang2019sample}): find a sufficient condition characterized by the so-called dual certificate and then construct the dual certificate. In fact, such routines are quite common in related sparse recovery problems such as compressed sensing and low-rank matrix recovery; see \cite{candes2006robust,candes2009exact}. But differently from previous kernel-based techniques for atomic norm methods, our approach is kernel-free and involves no frequency separation condition.

\subsection{Dual Certificate}
The following proposition is a corollary to the proof of \cite[Theorem 4]{yang2016exact}.

\begin{prop} Computing the weighted atomic norm by solving \eqref{eq:WAN} produces the true frequencies $\lbra{f_k}$ if there exists a vector-valued dual certificate $Q(f) = \m{a}^{H}(f)\m{V}$, $\m{V}\in\bC^{N\times L}$ satisfying that
{\lentwo\equa{Q\sbra{f_k}
&=& w(f_k) \m{\phi}_k,\quad k=1,\dots,K;  \label{eq:Qfk} \\
\twon{Q\sbra{f}}^2
&<& w^2(f), \quad f\in\bT \setminus \lbra{f_k}. \label{eq:Qf}}} \label{prop:dualcert}
\end{prop}

\begin{proof} Note that the weighted atomic norm in \eqref{eq:watomn} can be identified as a specialized atomic norm induced by the atomic set
\equ{\cA^w = \lbra{w^{-1}(f)\m{a}(f)\m{\phi}: \; f\in\bT,\; \m{\phi}\in\bC^{L\times1},\; \twon{\m{\phi}}=1}. \label{eq:atomset1}}
It then follows from the proof of \cite[Theorem 4]{yang2016exact} that computing such atomic norm produces the true frequencies $\lbra{f_k}$ if there exists a dual certificate $\widetilde Q(f) = w^{-1}(f)\m{a}^{H}(f)\m{V}$, $\m{V}\in\bC^{N\times L}$ satisfying that
{\lentwo\equa{\widetilde Q\sbra{f_k}
&=& \m{\phi}_k, \quad k=1,\dots,K;  \label{eq:wtQfk} \\
\twon{\widetilde Q\sbra{f}}^2
&<& 1, \quad f\in\bT \setminus \lbra{f_k}, \label{eq:wtQf}
}}or equivalently, if there exists $Q(f) = w(f)\widetilde Q(f) = \m{a}^{H}(f)\m{V}$ satisfying \eqref{eq:Qfk} and \eqref{eq:Qf}, completing the proof.
\end{proof}

The remaining task of the proof is to find such a dual certificate $Q(f) = \m{a}^{H}(f)\m{V}$ in Proposition \ref{prop:dualcert}.

\subsection{Construction of Dual Certificate}

We construct the dual certificate $Q(f) = \m{a}^{H}(f)\m{V}$ in Proposition \ref{prop:dualcert} by finding a proper $\m{V}$ in this subsection. It follows immediately from \eqref{eq:Qfk} that $\m{V}$ must satisfy that
\equ{Q\sbra{f_k} = \m{a}^{H}\sbra{f_k}\m{V} = w(f_k) \m{\phi}_k, \quad k=1,\dots,K,\label{eq:Qfk2}}
or equivalently,
\equ{\m{A}^{H}\m{V} = \diag\sbra{\m{\alpha}}\m{\Phi}, \label{eq:Qfk1}}
where $\m{A}$ is short for $\m{A}(\m{f})$, $\m{\Phi}$ is a $K\times L$ matrix of which the $k$th row is given by $\m{\phi}_k$, and $\m{\alpha} = \mbra{w\sbra{f_1},\dots,w\sbra{f_K}}^T$. To make \eqref{eq:Qf} hold true, a necessary condition is the following:
\equ{\frac{\text{d}}{\text{d}f} \sbra{\twon{Q\sbra{f_k}}^2 - w^2(f_k)}= 0, \quad k=1,\dots,K, \label{eq:der0}}
or equivalently,
\equ{\Re\lbra{Q'\sbra{f_k}Q^{H}\sbra{f_k} - \m{a}^{H}(f_k)\m{W}\m{a}'(f_k)} = 0,\label{eq:deriv0}}
where \eqref{eq:wf} was used.
Inserting \eqref{eq:Qfk} and \eqref{eq:Qfk2} into \eqref{eq:deriv0} yields that
\equ{\Re\lbra{\m{a}'^{H}\sbra{f_k}\m{V}\m{\phi}_k^{H} w(f_k)- \m{a}^{H}(f_k)\m{W}\m{a}'(f_k)} = 0.\label{eq:deriv1}}
Consequently, one feasible choice making \eqref{eq:deriv1} [and thus \eqref{eq:der0}] hold is to impose that
\equ{\m{a}'^{H}\sbra{f_k}\m{V}\m{\phi}_k^{H} = \frac{\m{a}^{H}(f_k)\m{W}\m{a}'(f_k)}{w(f_k)}, \quad k=1,\dots,K, \label{eq:Qdfk}}
or equivalently,
\equ{\diag\sbra{\m{A}'^{H}\m{V}\m{\Phi}^{H}}  = \m{\beta}, \label{eq:diagV}}
where $\m{A}' = \mbra{\m{a}'\sbra{f_1},\dots,\m{a}'\sbra{f_K}}$ and $\m{\beta}$ is a $K\times 1$ vector with the $k$th entry being $\frac{\boldsymbol{a}^{H}\left(f_{k}\right) \boldsymbol{W} \boldsymbol{a}^{\prime}\left(f_{k}\right)}{w\left(f_{k}\right)}$.

Note that \eqref{eq:Qfk1} and \eqref{eq:diagV} form a system of linear equations with respect to $\m{V}$ that can be written equivalently as:
\equ{ \begin{bmatrix}\m{I}_L\otimes \m{A}^{H}\\ \sbra{\m{\Phi}^T * \m{A}'}^{H} \end{bmatrix} \text{vec}\sbra{\m{V}} = \begin{bmatrix} \sbra{\m{\Phi}^T*\m{I}_K} \m{\alpha}  \\ \m{\beta} \end{bmatrix}.  \label{eq:Vsys2}}
We are going to show that the coefficient matrix in \eqref{eq:Vsys2} has full row rank and thus \eqref{eq:Vsys2} admits a solution. In order to do that, we need a few lemmas. The following result is a consequence of \cite[Proposition 1.1]{ellis1992factorization}.
\begin{lem} For every $k=1,\dots,K$, the $K+1$ vectors $\m{a}'(f_k), \m{a}(f_1), \dots, \m{a}(f_K)$ are linearly independent given $K<N$. \label{lem:linind}
\end{lem}

The following lemma presents results on the Kronecker, Khatri-Rao and Hadamard products of matrices that can be found in \cite{van2000ubiquitous,liu2008hadamard}.
\begin{lem} Let $\m{A}$, $\m{B}$, $\m{C}$, $\m{D}$ be matrices of proper dimensions from instance to instance, and additionally $\m{A}$, $\m{B}$ nonsingular in \eqref{eq:Kinv}. Then,
\lentwo{\equa{ \text{vec}\sbra{\m{A}\m{B}\m{C}}
&=& \sbra{\m{C}^T\otimes \m{A}} \text{vec}\sbra{\m{B}}, \label{eq:vec} \\ \sbra{\m{A}\otimes\m{B}}^{H}
&=& \m{A}^{H} \otimes \m{B}^{H}, \label{eq:KH} \\ \sbra{\m{A}\otimes\m{B}}^{-1}
&=& \m{A}^{-1} \otimes \m{B}^{-1}, \label{eq:Kinv} \\ \sbra{\m{A}\otimes \m{B}} \sbra{\m{C}\otimes \m{D}}
&=& \m{A}\m{C} \otimes \m{B}\m{D}, \label{eq:KK} \\ \sbra{\m{C}\otimes \m{D}} \sbra{\m{A}* \m{B}}
&=& \m{C}\m{A} * \m{D}\m{B}, \label{eq:KKR} \\ \sbra{\m{A} * \m{B}}^{H} \sbra{\m{A}* \m{B}}
&=& \m{A}^{H}\m{A} \odot \m{B}^{H}\m{B}. \label{eq:KRKR}
}} \label{lem:prodprop}
\end{lem}

The following result is part of the Schur product theorem; see \cite{schur1911bemerkungen} and \cite[Theorem 18]{yang2022nonasymptotic}.
\begin{lem} If $\m{A}$ is positive definite and $\m{B}$ is positive semidefinite with a positive diagonal, then $\m{A}\odot \m{B}$ is positive definite. \label{lem:GSPT}
\end{lem}

Now we are ready to show the following result.

\begin{lem} If $\m{\Phi}$ has full row rank and the frequencies $\lbra{f_k}$ are distinct, then the $\sbra{L+1}K \times LN$ coefficient matrix $\begin{bmatrix}\m{I}_L\otimes \m{A}^{H} \\ \sbra{\m{\Phi}^T * \m{A}'}^{H} \end{bmatrix}$ has full row rank. \label{lem:SysMatinv}
\end{lem}

\begin{proof} It suffices to show that the Gram matrix
\equ{\begin{split}\begin{bmatrix}\m{I}_L\otimes \m{A}^{H} \\ \sbra{\m{\Phi}^T * \m{A}'}^{H} \end{bmatrix}\begin{bmatrix}\m{I}_L\otimes \m{A}^{H} \\ \sbra{\m{\Phi}^T * \m{A}'}^{H} \end{bmatrix}^H
&= \begin{bmatrix}\m{I}_L\otimes \m{A}^{H} \\ \sbra{\m{\Phi}^T * \m{A}'}^{H} \end{bmatrix} \begin{bmatrix}\m{I}_L\otimes \m{A} & \m{\Phi}^T * \m{A}' \end{bmatrix} \\
&= \begin{bmatrix} \m{I}_L\otimes \m{A}^{H}\m{A} & \m{\Phi}^T*\m{A}^{H}\m{A}' \\ \sbra{\m{\Phi}^T*\m{A}^{H}\m{A}'}^{H} & \overline{\m{\Phi}}{\m{\Phi}}^T\odot \m{A}'^{H}\m{A}' \end{bmatrix}\end{split}}
is positive definite, where we have used \eqref{eq:KH}, \eqref{eq:KK}, \eqref{eq:KKR} and \eqref{eq:KRKR} of Lemma \ref{lem:prodprop}. Since $\m{I}_L\otimes \m{A}^{H}\m{A}$ is positive definite, we need only to show that its Schur complement is also positive definite which is given by
\equ{\begin{split}
&\overline{\m{\Phi}}{\m{\Phi}}^T\odot \m{A}'^{H}\m{A}' - \sbra{\m{\Phi}^T*\m{A}^{H}\m{A}'}^{H} \mbra{\m{I}_L\otimes \m{A}^{H}\m{A}}^{-1} \sbra{\m{\Phi}^T*\m{A}^{H}\m{A}'}\\
&=\overline{\m{\Phi}}{\m{\Phi}}^T\odot \m{A}'^{H}\m{A}' - \sbra{\m{\Phi}^T*\m{A}^{H}\m{A}'}^{H} \mbra{\m{I}_L\otimes \sbra{\m{A}^{H}\m{A}}^{-1}} \sbra{\m{\Phi}^T*\m{A}^{H}\m{A}'}\\
&=\overline{\m{\Phi}}{\m{\Phi}}^T\odot \m{A}'^{H}\m{A}' - \sbra{\m{\Phi}^T*\m{A}^{H}\m{A}'}^{H} \mbra{\m{I}_L\otimes \sbra{\m{A}^{H}\m{A}}^{-\frac{1}{2}}}^2 \sbra{\m{\Phi}^T*\m{A}^{H}\m{A}'}\\
&=\overline{\m{\Phi}}{\m{\Phi}}^T\odot \m{A}'^{H}\m{A}' - \sbra{\m{\Phi}^T*\sbra{\m{A}^{H}\m{A}}^{-\frac{1}{2}}\m{A}^{H}\m{A}'}^{H} \sbra{\m{\Phi}^T*\sbra{\m{A}^{H}\m{A}}^{-\frac{1}{2}}\m{A}^{H}\m{A}'}\\
&=\overline{\m{\Phi}}{\m{\Phi}}^T\odot \m{A}'^{H}\m{A}' - \overline{\m{\Phi}}{\m{\Phi}}^T\odot \m{A}'^{H}\m{A}\sbra{\m{A}^{H}\m{A}}^{-1}\m{A}^{H} \m{A}'\\
&=\overline{\m{\Phi}}{\m{\Phi}}^T\odot \m{A}'^{H}\mbra{\m{I}_N-\m{A}\sbra{\m{A}^{H}\m{A}}^{-1}\m{A}^{H}} \m{A}', \end{split} \label{eq:schurcomp}}
where we have consecutively applied \eqref{eq:Kinv}, \eqref{eq:KK}, \eqref{eq:KKR} and \eqref{eq:KRKR} in the first four equalities.

We next study the two matrix factors in the Hadamard product in \eqref{eq:schurcomp}. The first factor $\overline{\m{\Phi}}{\m{\Phi}}^T = \overline{\m{\Phi}{\m{\Phi}}^H}$ is positive definite since $\m{S}$, and thus $\m{\Phi}$, has full row rank by assumption. The proof will be completed by using Lemma \ref{lem:GSPT} if we can show that the second factor $\m{A}'^{H}\mbra{\m{I}_N-\m{A}\sbra{\m{A}^{H}\m{A}}^{-1}\m{A}^{H}} \m{A}'$ is positive semidefinite and has a positive diagonal. To this end, let $\m{A} = \begin{bmatrix}\m{U} & \m{U}_{\perp} \end{bmatrix} \begin{bmatrix} \m{\Lambda} \\ \m{0} \end{bmatrix} \m{V}^H = \m{U}\m{\Lambda}\m{V}^H$ be the singular value decomposition (SVD) of $\m{A}$. It is easy to verify that the second factor
\equ{\m{A}'^{H}\mbra{\m{I}_N-\m{A}\sbra{\m{A}^{H}\m{A}}^{-1}\m{A}^{H}}\m{A}' = \m{A}'^{H}\m{U}_{\perp}\m{U}_{\perp}^H\m{A}'}
is positive semidefinite. Now suppose its $(k,k)$ entry, which is given by
\equ{\m{a}'^H(f_k) \m{U}_{\perp}\m{U}_{\perp}^H\m{a}'(f_k) = \twon{\m{U}_{\perp}^H\m{a}'(f_k)}^2, \label{eq:Uperpad}}
is zero. We have immediately that $\m{U}_{\perp}^H\m{a}'(f_k)$ is a zero vector and thus, $\m{a}'(f_k)$ belongs to the range space of $\m{U}$ that is also the range space of $\m{A}$. This means that $\m{a}'(f_k)$ and the $K$ columns of $\m{A}$ are linearly dependent, which cannot be true according to Lemma \ref{lem:linind} and thus leads to contradiction. So we have proved that the second factor is positive semidefinite and has a positive diagonal, completing the proof.
\end{proof}

We are ready to construct the dual certificate $Q(f) = \m{a}^H(f)\m{V}$ by defining $\m{V}$ such that
\equ{ \text{vec}\sbra{\m{V}} = \begin{bmatrix}\m{I}_L\otimes \m{A}^{H} \\ \sbra{\m{\Phi}^T * \m{A}'}^{H} \end{bmatrix}^{\dag} \begin{bmatrix} \sbra{\m{\Phi}^T*\m{I}_K} \m{\alpha}  \\ \m{\beta} \end{bmatrix}.  \label{eq:V}}
It follows immediately from Lemma \ref{lem:SysMatinv} that the $\m{V}$ defined above is a solution to \eqref{eq:Vsys2}. Consequently, the constructed dual certificate $Q(f) = \m{a}^H(f)\m{V}$ satisfies \eqref{eq:Qfk} and \eqref{eq:der0}. To complete the proof, it suffices to verify that $Q(f) = \m{a}^H(f)\m{V}$ also satisfies \eqref{eq:Qf}, which is the task of the ensuing subsection.

\subsection{Validation of Dual Certificate}
We verify that $Q(f) = \m{a}^H(f)\m{V}$ satisfies \eqref{eq:Qf} under the assumptions of Theorem \ref{thm:noiseless} in this subsection, which completes the proof. We need the following proposition.

\begin{lem}
\equ{\twon{Q^{(l)}(f)} \leq C_l \max\lbra{\epsilon^{\frac{1}{2}},\,\epsilon}, \quad l=0,1,2,\label{eq:Qlbound}}
where $Q^{(l)}(f)$ denotes the $l$th derivative of $Q(f)$, with $Q^{(0)}(f)=Q(f)$, and $\lbra{C_l}$ are constants independent of $\epsilon$. \label{prop:Qbound}
\end{lem}

\begin{proof}
Note that $Q^{(l)}(f) = \m{a}^{(l)H}(f)\m{V}$ and thus, for $l=0, 1, 2$,
\equ{\twon{Q^{(l)}(f)} \leq \twon{\m{a}^{(l)H}(f)}\twon{\m{V}} = \twon{\m{a}^{(l)}(f)}\twon{\m{V}}. \label{eq:Qlupp}}
It is easy to see that $\twon{\m{a}^{(l)}(f)}$ is a constant depending only on $l$. Consequently, it suffices to bound $\twon{\m{V}}$ with a constant times $\max\lbra{\epsilon^{\frac{1}{2}},\,\epsilon}$. By using \eqref{eq:V}, we have that
\equ{\begin{split}\twon{\m{V}}
&\leq \frobn{\m{V}}\\
&= \twon{\text{vec}\sbra{\m{V}}} \\
&\leq \twon{\begin{bmatrix}\m{I}_L\otimes \m{A}^{H} \\ \sbra{\m{\Phi}^T * \m{A}'}^{H} \end{bmatrix}^{\dag}} \twon{\begin{bmatrix} \sbra{\m{\Phi}^T*\m{I}_K} \m{\alpha}  \\ \m{\beta} \end{bmatrix}}\\
&\leq \twon{\begin{bmatrix}\m{I}_L\otimes \m{A}^{H} \\ \sbra{\m{\Phi}^T * \m{A}'}^{H} \end{bmatrix}^{\dag}} \sbra{\twon{\sbra{\m{\Phi}^T*\m{I}_K} \m{\alpha}} + \twon{\m{\beta}}} \\
&\leq \twon{\begin{bmatrix}\m{I}_L\otimes \m{A}^{H} \\ \sbra{\m{\Phi}^T * \m{A}'}^{H} \end{bmatrix}^{\dag}} \sbra{\twon{\m{\Phi}^T*\m{I}_K} \twon{\m{\alpha}} + \twon{\m{\beta}}}.  \end{split} \label{eq:boundV}}
To complete the proof, therefore, we need only to show that both $\lbra{\alpha_k}$ and $\lbra{\beta_k}$ can be bounded from above by a constant times $\max\lbra{\epsilon^{\frac{1}{2}},\,\epsilon}$.

We first do some preparations. Recall that
$\widehat{\m{R}} = \frac{1}{L} \m{Y}\m{Y}^H = \m{A}\sbra{\frac{1}{L}\m{S}\m{S}^H}\m{A}^H$
and let
\equ{\widehat{\m{R}} = \begin{bmatrix}\m{U} & \m{U}_{\perp}\end{bmatrix} \begin{bmatrix}\m{\Lambda}^2 & \m{0} \\ \m{0} & \m{0} \end{bmatrix} \begin{bmatrix}\m{U} & \m{U}_{\perp}\end{bmatrix}^{H} = \m{U}\m{\Lambda}^2\m{U}^H}
be its eigen-decomposition, where $\m{\Lambda}^2 = \diag\sbra{\lambda_1,\dots,\lambda_K}$ with $\lambda_1\geq \dots \geq \lambda_K>0$. It is easy to see that $\m{A}$ and $\m{U}$ share the same range space and thus,
{\lentwo\equa{\m{U}^{H}\m{a}\sbra{f_k}
&=& \m{a}\sbra{f_k}, \quad k=1,\dots,K,\label{eq:Ua} \\ \m{U}_{\perp}^{H}\m{a}\sbra{f_k}
&=& \m{0}, \quad k=1,\dots,K, \label{eq:Uperpa1} \\ \m{U}_{\perp}^{H}\m{a}\sbra{f}
&\neq& \m{0},\quad f\in\bT\setminus \lbra{f_k}. \label{eq:Uperpa2}
}}Moreover,
\equ{\begin{split}\m{W}
&= \sbra{\m{I} + \epsilon^{-1} \widehat{\m{R}}}^{-1}\\
&= \sbra{\m{I} + \epsilon^{-1} \begin{bmatrix}\m{U} & \m{U}_{\perp}\end{bmatrix} \begin{bmatrix}\m{\Lambda}^2 & \m{0} \\ \m{0} & \m{0} \end{bmatrix} \begin{bmatrix}\m{U} & \m{U}_{\perp}\end{bmatrix}^{H} }^{-1}\\
&= \begin{bmatrix}\m{U} & \m{U}_{\perp}\end{bmatrix} \begin{bmatrix}\sbra{\m{I} + \epsilon^{-1} \m{\Lambda}^2}^{-1} & \m{0} \\ \m{0} & \m{I} \end{bmatrix} \begin{bmatrix}\m{U} & \m{U}_{\perp}\end{bmatrix}^{H}\\
&= \m{U}\diag\sbra{\frac{\epsilon}{\lambda_1+\epsilon},\dots, \frac{\epsilon}{\lambda_K+\epsilon}} \m{U}^{H} + \m{U}_{\perp}\m{U}_{\perp}^{H}. \end{split} \label{eq:Winepsilon}}

Now we are ready to bound $\lbra{\alpha_k = w\sbra{f_k}}$. Making use of \eqref{eq:Ua}, \eqref{eq:Uperpa1} and \eqref{eq:Winepsilon}, we have for every $k=1,\dots,K$ that
\equ{\begin{split}\alpha_k^2
&= w^2(f_k)\\
&= \m{a}^{H}\sbra{f_k}\m{W}\m{a}\sbra{f_k}\\
&= \m{a}^{H}\sbra{f_k}\m{U}\diag\sbra{\frac{\epsilon}{\lambda_1+\epsilon},\dots, \frac{\epsilon}{\lambda_K+\epsilon}} \m{U}^{H}\m{a}\sbra{f_k} + \m{a}^H\sbra{f_k}\m{U}_{\perp}\m{U}_{\perp}^{H}\m{a}\sbra{f_k}\\
&= \m{a}^{H}\sbra{f_k}\m{U}\diag\sbra{\frac{\epsilon}{\lambda_1+\epsilon},\dots, \frac{\epsilon}{\lambda_K+\epsilon}} \m{U}^{H}\m{a}\sbra{f_k}\\
&= \m{a}^{H}\sbra{f_k}\diag\sbra{\frac{\epsilon}{\lambda_1+\epsilon},\dots, \frac{\epsilon}{\lambda_K+\epsilon}} \m{a}\sbra{f_k}. \end{split}}
It follows immediately that
\equ{\frac{N\epsilon}{\lambda_1+\epsilon} = \frac{\epsilon}{\lambda_1+\epsilon} \twon{\m{a}\sbra{f_k}}^2 \leq \alpha_k^2 \leq \frac{\epsilon}{\lambda_K} \twon{\m{a}\sbra{f_k}}^2 = \frac{N\epsilon}{\lambda_K} }
and thus,
\equ{\sqrt{\frac{N\epsilon}{\lambda_1+\epsilon}} \leq \alpha_k \leq \sqrt{\frac{N}{\lambda_K}}\epsilon^{\frac{1}{2}}. \label{eq:wfkorder}}
The upper bound above is what we need.

%\equ{\begin{split}\alpha_k^2
%&= w^2(f_k)\\
%&= \m{a}^{H}\sbra{f_k}\m{W}\m{a}\sbra{f_k}\\
%&= \m{a}^{H}\sbra{f_k}\m{U}\diag\sbra{\frac{\epsilon}{\lambda_1+\epsilon},\dots, \frac{\epsilon}{\lambda_K+\epsilon}} \m{U}^{H}\m{a}\sbra{f_k} + \m{a}^H\sbra{f_k}\m{U}_{\perp}\m{U}_{\perp}^{H}\m{a}\sbra{f_k}\\
%&= \m{a}^{H}\sbra{f_k}\m{U}\diag\sbra{\frac{\epsilon}{\lambda_1+\epsilon},\dots, \frac{\epsilon}{\lambda_K+\epsilon}} \m{U}^{H}\m{a}\sbra{f_k}\\
%&\leq \twon{\m{a}\sbra{f_k}}^2 \cdot\twon{\m{U}}^2 \cdot \twon{\diag\sbra{\frac{\epsilon}{\lambda_1+\epsilon},\dots, \frac{\epsilon}{\lambda_K+\epsilon}}}\\
%&= N \cdot \frac{\epsilon}{\lambda_K+\epsilon}\\
%&\leq \frac{N}{\lambda_K}\epsilon. \end{split} \label{w2fk}}
%Therefore,
%\equ{\alpha_k \leq \sqrt{\frac{N}{\lambda_K}}\epsilon^{\frac{1}{2}},}
%as required.

We next bound $\lbra{\beta_k = \frac{\boldsymbol{a}^{H}\left(f_{k}\right) \boldsymbol{W} \boldsymbol{a}^{\prime}\left(f_{k}\right)}{\alpha_k}}$. It follows from \eqref{eq:Winepsilon} and \eqref{eq:Uperpa1} that for every $k=1,\dots,K$,
\equ{\begin{split}\abs{\m{a}^{H}(f_k)\m{W}\m{a}'(f_k)}
&= \abs{\m{a}^{H}\sbra{f_k}\m{U}\diag\sbra{\frac{\epsilon}{\lambda_1+\epsilon},\dots, \frac{\epsilon}{\lambda_K+\epsilon}} \m{U}^H\m{a}'(f_k)} \\
&\leq \twon{\m{a}\sbra{f_k}} \twon{\m{U}}^2 \cdot \frac{\epsilon}{\lambda_K} \cdot \twon{\m{a}'(f_k)}\\
&= \sqrt{N} \cdot\frac{\epsilon}{\lambda_K} \cdot2\pi \sqrt{\frac{(N-1)N(2N-1)}{6}}\\
&\leq \frac{2\pi N^2\epsilon}{\sqrt{3}\lambda_K}, \end{split} \label{eq:vfkorder}}
where the second equality uses $\twon{\m{a}\sbra{f_k}} = \sqrt{N}$, $\twon{\m{U}}^2=1$ and \equ{\twon{\m{a}'(f_k)} = \twon{2\pi i\cdot \diag\sbra{0,\dots,N-1}\m{a}(f_k)} = 2\pi \sqrt{\frac{(N-1)N(2N-1)}{6}}.}
Making use of \eqref{eq:wfkorder} and \eqref{eq:vfkorder}, we have that for every $k=1,\dots,K$,
\equ{\abs{\beta_k} = \frac{\abs{\m{a}^{H}(f_k)\m{W}\m{a}'(f_k)}}{\alpha_k} \leq \frac{2\pi N^2\epsilon}{\sqrt{3}\lambda_K} \cdot \sqrt{\frac{\lambda_1+\epsilon}{N\epsilon}} \leq \frac{2\pi N^{\frac{3}{2}}}{\sqrt{3}\lambda_K}\cdot \sqrt{\lambda_1\epsilon+\epsilon^2} \leq \frac{2\pi N^{\frac{3}{2}}}{\lambda_K}\max\lbra{\sqrt{\lambda_1}\epsilon^{\frac{1}{2}},\epsilon}. \label{eq:betaupp}}
Inserting \eqref{eq:wfkorder} and \eqref{eq:betaupp} into \eqref{eq:boundV} and then inserting \eqref{eq:boundV} into \eqref{eq:Qlupp} complete the proof.
\end{proof}

Lemma \ref{prop:Qbound} and its proof will be used to prove the following two propositions.

\begin{prop} There exist $\nu_0>0$ and $\epsilon'_0>0$ such that for every $k=1,\dots,K$, if $0<\epsilon\leq\epsilon'_0$, then
\equ{\twon{Q\sbra{f}}^2 < w^2(f), \quad f\in [f_k-\nu_0,\; f_k) \cup (f_k,\; f_k+\nu_0].} \label{prop:near}
\end{prop}
\begin{proof} Consider the function
\equ{g_{\epsilon}(f) = w^2(f) - \twon{Q\sbra{f}}^2 \label{eq:g}}
of $f$ with $\epsilon>0$.
Since \eqref{eq:Qfk} and \eqref{eq:der0} are satisfied, we have immediately that
\equ{g_{\epsilon}(f_k) = g'_{\epsilon}(f_k) = 0,\quad k=1,\dots,K.}
To prove the lemma, therefore, it suffices to show that $g_{\epsilon}(f)$ is a strictly convex function of $f$ on $\mbra{f_k-\nu_0,\;f_k+\nu_0}$ for every $k=1,\dots,K$ and $0<\epsilon\leq\epsilon'_0$.

In order to do that, we first argue that it suffices to prove that for every $k=1,\dots,K$,
\equ{\lim_{\epsilon\rightarrow0} g''_{\epsilon}(f_k) > 0.  \label{eq:Qfmw22d}}
In particular, note that once \eqref{eq:Qfmw22d} is shown to be true, according to the continuity of $g''_{\epsilon}(f)$ as a function of $\sbra{f,\epsilon}\in \bT \times (0,+\infty)$, there will exist $\nu_k,\epsilon'_k>0$ for every $k=1,\dots,K$ such that
\equ{g''_{\epsilon}(f) \geq \frac{1}{2}\lim_{\epsilon\rightarrow0} g''_{\epsilon}(f_k)>0, \quad \sbra{f, \epsilon} \in \mbra{f_k-\nu_k, f_k+\nu_k}\times (0,\epsilon'_k].}
This means that for every $\epsilon \leq \epsilon'_k$, $g_{\epsilon}(f)$ is a strictly convex function of $f$ on $\mbra{f_k-\nu_k, f_k+\nu_k}$. The proof is completed by letting $\epsilon'_0 = \min \lbra{\epsilon'_1,\dots,\epsilon'_K}$ and $\nu_0 = \min \lbra{\nu_1,\dots,\nu_K}$.

The rest of the proof is devoted to showing \eqref{eq:Qfmw22d}. Since
\equ{\frac{\text{d}^2}{\text{d}f^2} \twon{Q\sbra{f_k}}^2 = 2\twon{Q'(f_k)}^2 + 2\Re\lbra{Q''(f_k)Q^{H}(f_k)}, \label{eq:Q22dfk}}
it follows from Lemma \ref{prop:Qbound} that
\equ{\lim_{\epsilon\rightarrow0} \frac{\text{d}^2}{\text{d}f^2} \twon{Q\sbra{f_k}}^2 = 0. \label{eq:Qfk2dgo0}}
Moreover, recall \eqref{eq:wf} and \eqref{eq:Winepsilon} and we have
\equ{\begin{split}\lim_{\epsilon\rightarrow0}\frac{\text{d}^2}{\text{d}f^2} w^2(f_k)
&= \lim_{\epsilon\rightarrow0} 2\m{a}'^{H}(f_k)\m{W}\m{a}'(f_k) + 2\Re\lbra{\m{a}^{H}(f_k)\m{W}\m{a}''(f_k)} \\
&= \lim_{\epsilon\rightarrow0} 2\m{a}'^{H}(f_k)\m{W}\m{a}'(f_k)\\
&= \lim_{\epsilon\rightarrow0} 2\m{a}'^{H}(f_k)\mbra{\m{U}\diag\sbra{\frac{\epsilon}{\lambda_1+\epsilon},\dots, \frac{\epsilon}{\lambda_K+\epsilon}} \m{U}^{H} + \m{U}_{\perp}\m{U}_{\perp}^{H}} \m{a}'(f_k)\\
&= 2\m{a}'^{H}(f_k)\m{U}_{\perp} \m{U}_{\perp}^{H}\m{a}'(f_k)\\
&= 2\twon{\m{U}_{\perp}^{H}\m{a}'(f_k)}^2\\
&> 0, \end{split} \label{eq:w2fk2d}}
where the second equality holds since
\equ{\lim_{\epsilon\rightarrow0} \m{a}^{H}(f_k)\m{W}\m{a}''(f_k) = 0,}
which can be shown similarly to \eqref{eq:vfkorder}, and the last inequality comes from the arguments around \eqref{eq:Uperpad}.
Combining \eqref{eq:Qfk2dgo0} and \eqref{eq:w2fk2d} yields that
\equ{\lim_{\epsilon\rightarrow0} g''_{\epsilon}(f_k) = \lim_{\epsilon\rightarrow0}\frac{\text{d}^2}{\text{d}f^2} w^2(f_k) - \lim_{\epsilon\rightarrow0} \frac{\text{d}^2}{\text{d}f^2} \twon{Q\sbra{f_k}}^2 > 0,}
which concludes \eqref{eq:Qfmw22d} and completes the proof.
\end{proof}

\begin{prop} There exists $\epsilon''_0>0$ such that if $0<\epsilon\leq \epsilon''_0$, then
\equ{\twon{Q\sbra{f}}^2 < w^2(f), \quad f\in \bT \setminus \bigcup_{k=1}^K[f_k-\nu_0,\; f_k+\nu_0], \label{eq:Qlw2}}
where $\nu_0>0$ was defined in Proposition \ref{prop:near}. \label{prop:far}
\end{prop}
\begin{proof} Recall \eqref{eq:Winepsilon} and we have
\equ{w^2(f) = \m{a}^H(f)\m{W}\m{a}(f) \geq \m{a}^H(f)\m{U}_{\perp}\m{U}_{\perp}^{H} \m{a}(f) = \twon{\m{U}_{\perp}^{H}\m{a}\sbra{f}}^2. \label{eq:w2fUa}}
Let
\equ{\zeta = \inf\lbra{\twon{\m{U}_{\perp}^{H}\m{a}\sbra{f}}^2:\; f\in \bT \setminus  \bigcup_{k=1}^K[f_k-\nu_0,\; f_k+\nu_0]}. \label{eq:zeta}}
We must have $\zeta>0$ since, otherwise, $\m{U}_{\perp}^{H}\m{a}\sbra{f}=\m{0}$ for some $f\in \bT \setminus  \bigcup_{k=1}^K[f_k-\nu_0,\; f_k+\nu_0]$, which cannot be true according to  \eqref{eq:Uperpa2}.
It follows immediately that for every $\epsilon>0$,
\equ{w^2(f) \geq \zeta>0, \quad f \in \bT \setminus \bigcup_{k=1}^K[f_k-\nu_0,\; f_k+\nu_0]. \label{eq:w2fgzeta}}
On the other hand, by Lemma \ref{prop:Qbound}, there exists $\epsilon''_0>0$ such that if $0<\epsilon \leq \epsilon''_0$, then
\equ{\twon{Q\sbra{f}}^2 \leq \frac{1}{2}\zeta, \quad f\in\bT. \label{eq:Qflhzeta}}
Combining \eqref{eq:w2fgzeta} and \eqref{eq:Qflhzeta} results in \eqref{eq:Qlw2}, completing the proof.
\end{proof}

Now we are ready to show that the constructed dual certificate $Q(f)$ satisfies \eqref{eq:Qf} under the assumption of Theorem \ref{thm:noiseless} and thus complete the whole proof. In fact, \eqref{eq:Qf} is a direct consequence of combining Proposition \ref{prop:near} and Proposition \ref{prop:far} under the assumption $0<\epsilon < \epsilon_0$ by defining $\epsilon_0 = \min\lbra{\epsilon'_0,\epsilon''_0}$.

\section{Numerical Results} \label{sec:simulation}

In this section we provide numerical results to validate our theoretical findings as well as to demonstrate the usefulness of the weighted atomic norm method in challenging scenarios with missing data and Gaussian noise.

\begin{figure}[t]
\centering
\includegraphics[width=10cm]{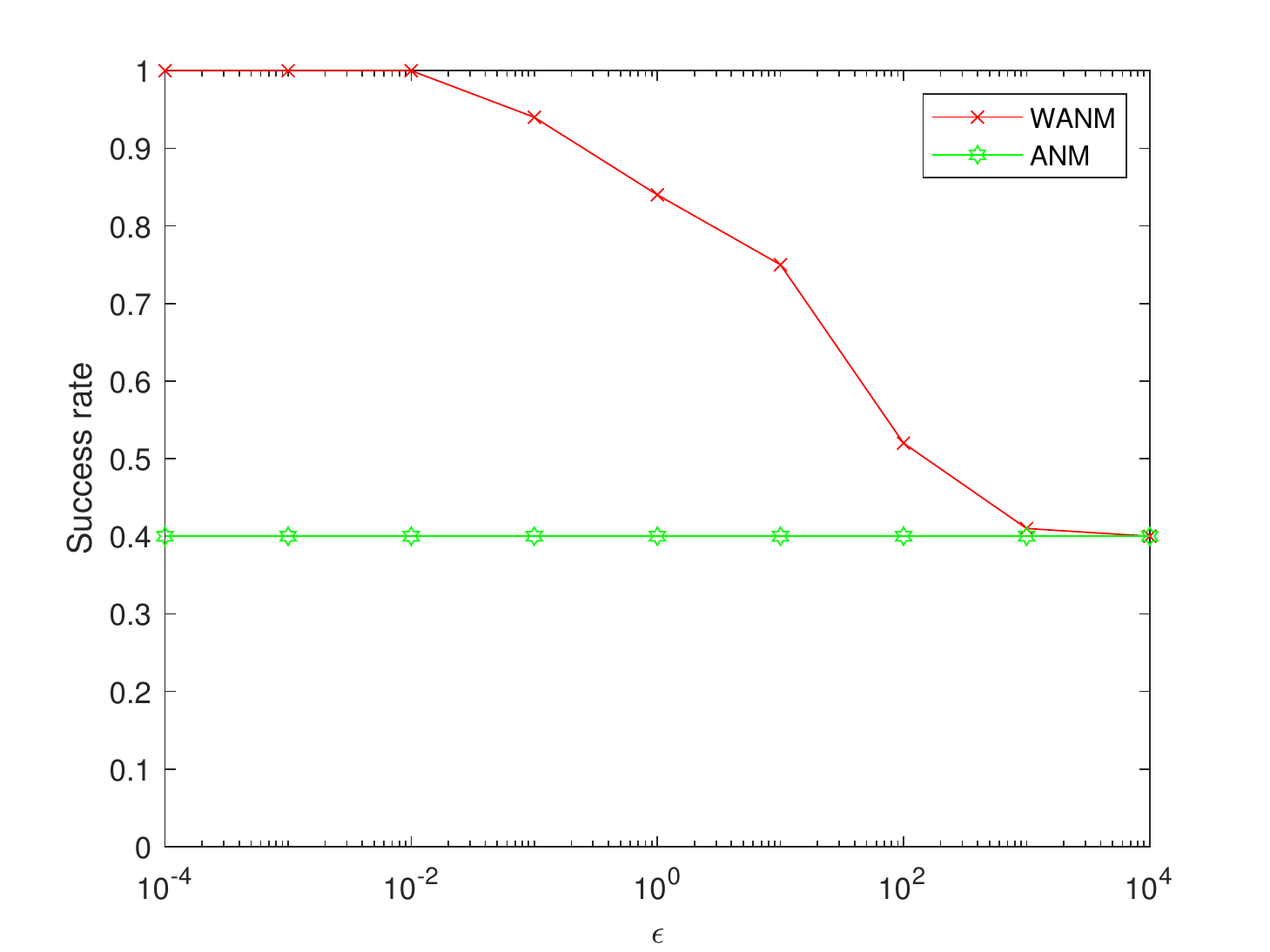}
\caption{Results of success rate of frequency recovery versus $\epsilon$, where $N=20$, $K=2$, $L=3$, $f_1= 0.3$ and $f_2 = f_1 + \frac{1}{2N}=0.325$.}
\label{fig:epsilon}
\end{figure}

We first consider the noiseless, full data case, as concerned in Theorem \ref{thm:noiseless}. In {\em Experiment 1}, we study the performance of the weighted atomic norm method (WANM) with respect to $\epsilon$. We set $N=20$, $K=2$, $L=3$, $f_1=0.3$ and $f_2 = 0.325$, implying that the frequency separation is $\frac{0.5}{N}=0.025$. The source signals in $\m{S}$ are generated from a standard complex Gaussian distribution. We consider $\epsilon = 10^{-4},10^{-3},\dots,10^4$ in WANM. We randomly generate 100 problems and solve them using WANM, in which we consider $\epsilon = 10^{-4},10^{-3},\dots,10^4$. We claim that the frequencies are successfully recovered if the root mean squared error of frequency recovery is below $10^{-4}$. The atomic norm method (ANM) is also used to solve the problems for comparison that does not depend on $\epsilon$. Our numerical results are presented in Fig.~\ref{fig:epsilon}. It is seen that the performance of WANM approaches that of ANM as $\epsilon$ grows very large, which is true since the weighting function in WANM becomes constant and WANM degenerates into ANM as $\epsilon\rightarrow +\infty$. As $\epsilon$ becomes small (below $10^{-2}$ in this case), $100\%$ success occurs for WANM, which validates Theorem \ref{thm:noiseless}.

\begin{figure}[t]
\centering
\includegraphics[width=10cm]{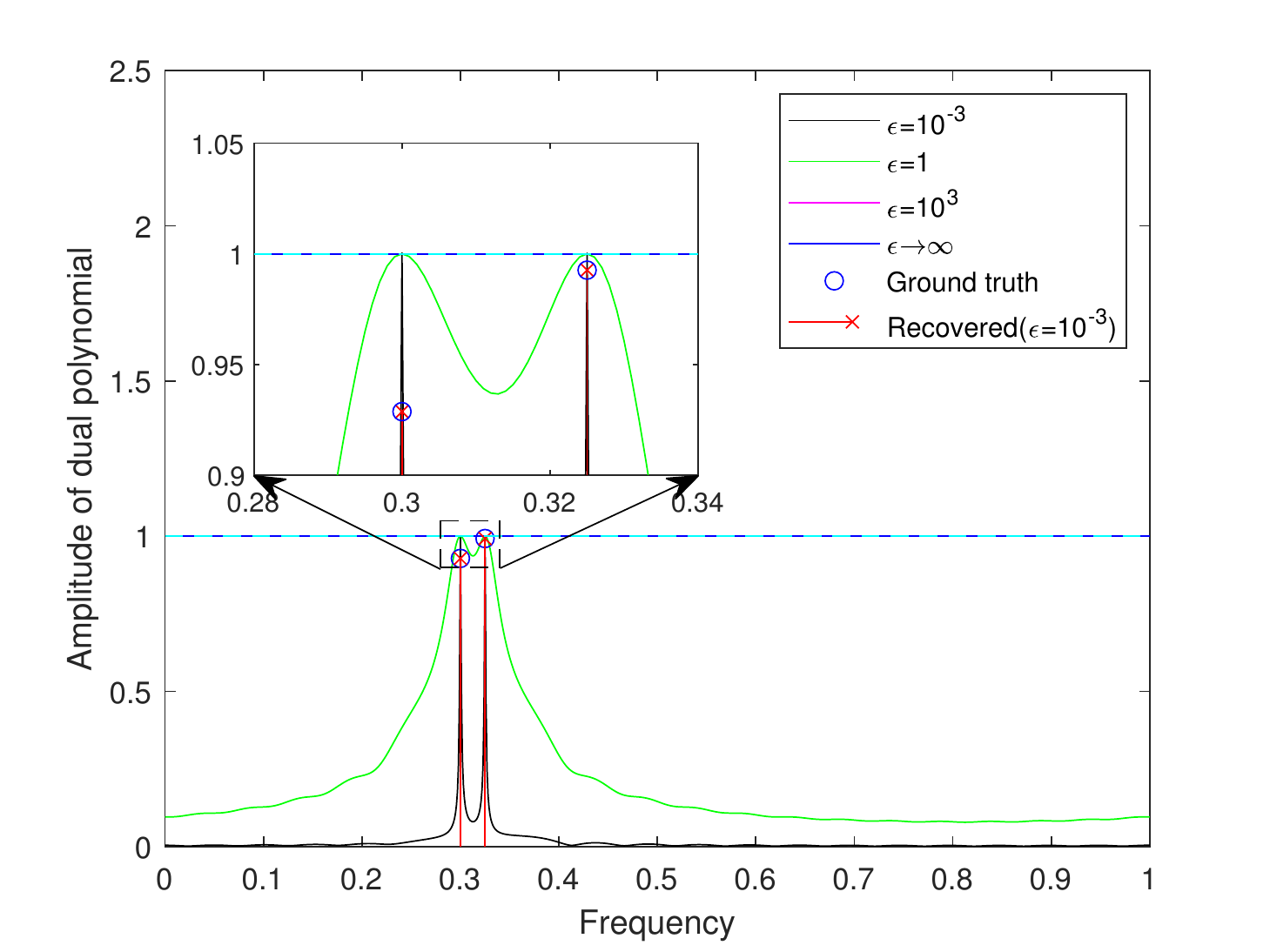}
\caption{Plots of the weighted dual polynomial $w^{-1}(f)Q(f)$ of WANM with respect to frequency $f$ for different values of $\epsilon$. WANM degenerates into ANM in the case $\epsilon\rightarrow \infty$.}
\label{fig:dualpoly}
\end{figure}

We present in Fig.~\ref{fig:dualpoly} the dual certificates of WANM and ANM for one problem. For the purpose of better illustration, we plot the amplitude of the scaled dual certificate $\widetilde Q(f) = w^{-1}(f)Q(f)$ with respect to $f$ for different values of $\epsilon$. It follows from \eqref{eq:wtQfk} and \eqref{eq:wtQf} that the frequency estimates of WANM (and ANM) can be identified from the points of tangency between the curve $\twon{\widetilde Q(f)}$ and the straight line at unit. It is seen that WANM succeeds to recover the frequencies when $\epsilon=10^{-3},1$, while it fails when $\epsilon=10^{3}$ and $\epsilon \rightarrow +\infty$ (the latter corresponds to ANM). The dual polynomial gets sharp around the true frequencies as $\epsilon$ becomes small. Interestingly, when $\epsilon$ is large, we have $\abs{\widetilde Q(f)}\equiv 1$, as shown in Fig.~\ref{fig:dualpoly}. It occurs because the solved Toeplitz matrix for WANM has full rank (see \eqref{eq:Vandec}) and in this case any value of frequency can be included in the Vandermonde decomposition (see \cite[Theorem 11.5]{yang2018sparse}).

In {\em Experiment 2}, we fix $\epsilon = 10^{-3}$ and vary the frequency separation $\Delta_f$ from $0$ to $\frac{2}{N}$ at a step of $\frac{0.05}{N}$ and use the other setups as in {\em Experiment 1}. Our numerical results are presented in Fig.~\ref{fig:epsilon}. It is seen that $100\%$ success occurs for ANM as $\Delta_f\geq \frac{0.85}{N}$, while this resolution limit is improved to $\frac{0.2}{N}$ for WANM.

\begin{figure}[t]
\centering
\includegraphics[width=10cm]{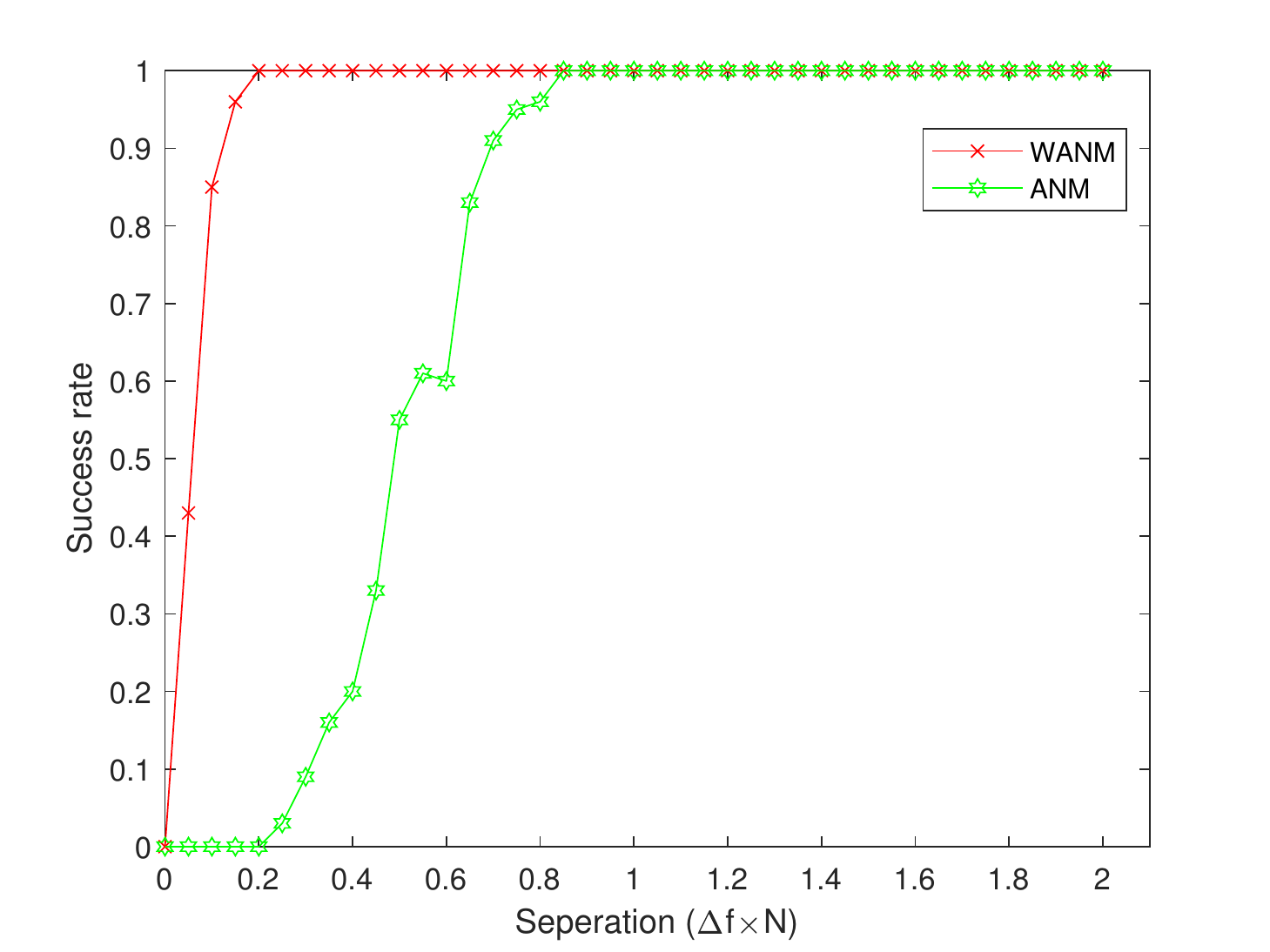}
\caption{Results of success rate of frequency recovery versus frequency separation $\Delta_f$, where $N=20$, $K=2$, $L=3$, $f_2 = f_1 + \Delta_f$ and $\epsilon=10^{-3}$.}
\label{fig:sepation}
\end{figure}

\begin{figure}[t]
\centering
\includegraphics[width=10cm]{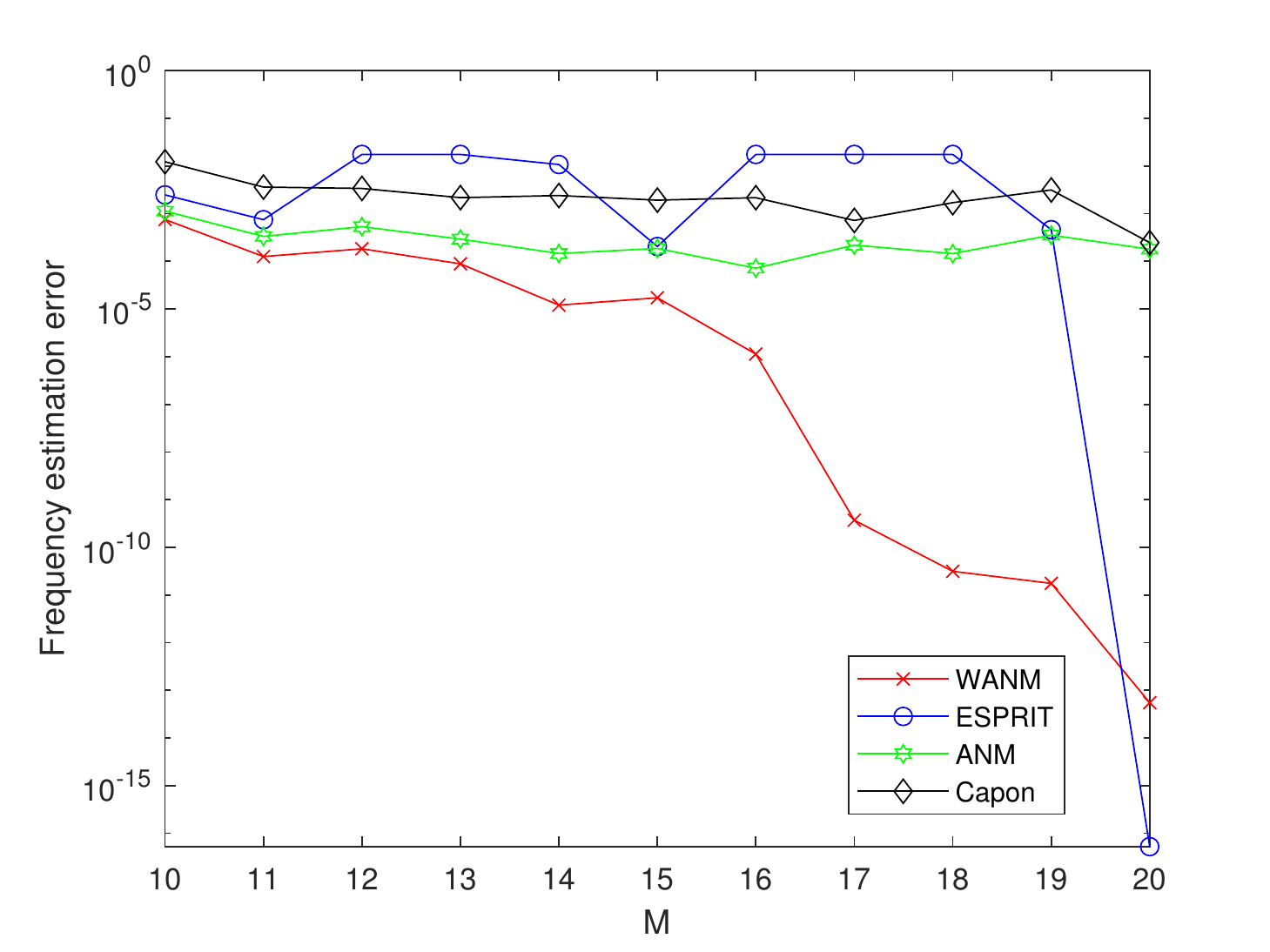}
\caption{Results of success rate of frequency recovery versus the number $M$ of observed rows of data matrix, where $N=20$, $K=2$, $L=3$, $\Delta_f=\frac{0.7}{N}$ and $\epsilon=10^{-3}$.}
\label{fig:missing}
\end{figure}

In {\em Experiment 3}, we study the missing data case in which only a subset of the rows of the data matrix $\m{Y}$ is observed, as in array signal processing. We randomly select $M$ out of $N$ rows from $\m{Y}$. We set $N=20$, $L=3$, $K=2$ and $\Delta f=\frac{0.7}{N}$ and use $\epsilon = 10^{-3}$ in WANM. ANM, ESPRIT and Capon's method are considered for comparison. Capon's method is implemented as in \eqref{eq:capon}, where $\epsilon$ is used for regularization in the case when $\widehat{\m{R}}$ is rank deficient. Consequently, WANM is a combination of Capon and ANM. Our results on the root mean squared error of frequency recovery with varying $M$ are presented in Fig.~\ref{fig:missing}. It is seen that WANM has the smallest error, followed by ANM. ESPRIT and Capon do not have satisfactory performance, except for $M=N$, since the data covariance matrix cannot be estimated accurately with missing data.

\begin{figure}[t]
\centering
\includegraphics[width=10cm]{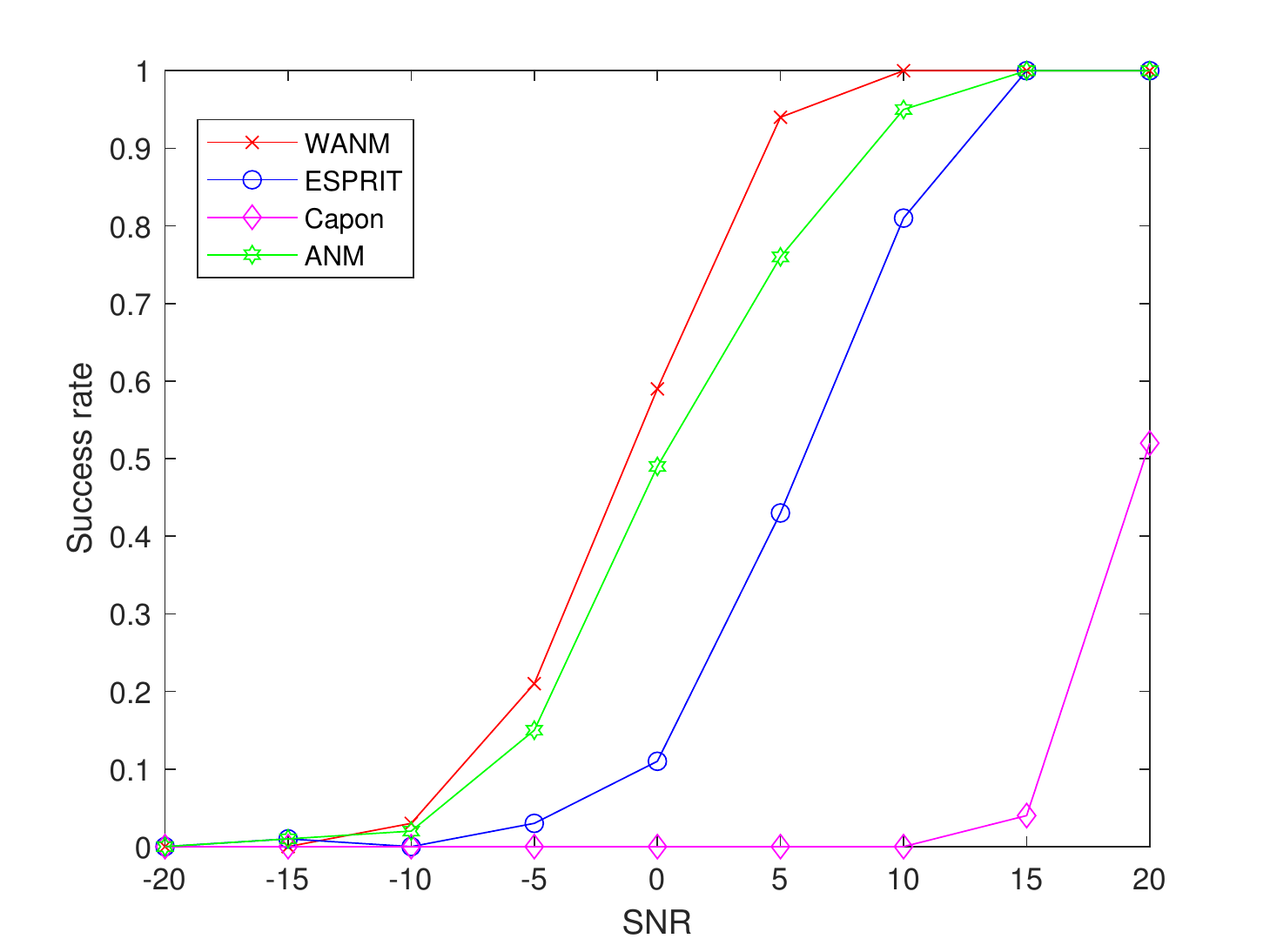}
\caption{Results of success rate versus SNR, where $N=20$, $K=2$, $L=10$, $\Delta_f=\frac{0.3}{N}$ and $\epsilon=10^{-3}$.}
\label{fig:noisy}
\end{figure}

In {\em Experiment 4}, we study the noisy data case in which the measurements are corrupted by additive white Gaussian noise. We set $N=20$, $K=2$, $L=10$ and $\Delta_f = \frac{0.3}{N}$. We set $\epsilon = 10^{-3}$ in WANM. We vary the signal-to-noise ratio (SNR) from $-20$dB to $20$dB. We say that the frequencies are successfully resolved if the absolute estimation error for each frequency is less than $\frac{1}{2}\Delta_f$. Our results on the success rate versus SNR is presented in Fig.~\ref{fig:noisy}. Again, it is seen that WANM has the best performance.

\section{Conclusion} \label{sec:conclusion}
In this paper, we provided a theoretical analysis showing that the resolution limit of the atomic norm method for spectral super-resolution can be overcome by including an adaptive weighting strategy. Numerical results are provided that corroborate our analysis and demonstrate the usefulness of the proposed method in practical scenarios with missing data and noise.

%studied the resolution limit of the atomic norm approaches in line spectral estimation. The resolution limit was overcome by adding a simple weighting scheme when niose free. Solving for a semipositive programming, we can accurately recover the frequency, provided that the frequency can be uniquely identified. A complete theoretical guarantee was given by constructing a dual certificate without kernel in the proof. Compared with conventional subspace methods, atomic norm approaches perform better under some worse conditions, but indeed have a minimum separation distance in the ideal condition. This paper compensates the gap between atomic norm approaches and conventional subspace methods in resolution. The simulations have shown that atomic norm approaches can exactly resolve frequencies in noiseless, and even in noise, these approaches have a higher resolution than other algorithms, intuitively. Generalization of this result to cases with coherent sources can be considered in the future.

%\bibliographystyle{IEEEtran}
%%\bibliography{myreferences1}
%\bibliography{C:/Users/nnd94/OneDrive/Research/MyWork/myreferences1}


%\cite{myreferences1}

\end{document}